\documentclass[onecolumn]{autart}

\usepackage[T1]{fontenc}
\usepackage[utf8]{inputenc}
\usepackage[USenglish]{babel}

\usepackage{amsmath}
\usepackage{amssymb}
\usepackage{mathtools}
\usepackage{commath}
\usepackage{bm}

\usepackage[caption=false,font=footnotesize]{subfig}

\usepackage[pdftex]{graphicx}
	\graphicspath{{plots/}{images/}}
	\DeclareGraphicsExtensions{.pdf,.jpeg,.png}
\usepackage{pgfplots}
	\pgfplotsset{compat=newest}
\usepackage{tikz}
\usepackage{tikzscale}
\usepackage{pdfpages}

\usepackage{xfrac}
\usepackage[per-mode=symbol,binary-units=true]{siunitx}
\usepackage[autostyle=false,english=american]{csquotes}
\usepackage{physics}
\usepackage[short,nocomma,c2]{packages/optidef}
\usepackage{algorithm}
\usepackage[noend]{algpseudocode}
\usepackage{hyperref}

\newlength{\figureheight}
\newlength{\figurewidth}

\def \Reals {\mathbb{R}}

\def \Nats {\mathbb{N}}

\def \Bins {\mathbb{B}}
\def \Trans {\top}

\DeclareMathAlphabet{\mymathbb}{U}{BOONDOX-ds}{m}{n}

\DeclareMathOperator{\diag}{diag}

\DeclareMathOperator{\Gauss}{\mathcal{N}}

\begin{document}

\begin{frontmatter}
	
\title{Trajectory Planning Under Environmental Uncertainty With Finite-Sample Safety Guarantees\thanksref{footnoteinfo}}
\thanks[footnoteinfo]{The work of M.~Kamgarpour is gratefully supported by the Swiss National Science Foundation, under the grant SNSF 200021\_172782, and by the NSERC Discovery Grant RGPAS\nobreakdash-2020\nobreakdash-00110. Corresponding author M.~Kamgarpour.}

\author[ETHZ_IfA]{Vasileios Lefkopoulos}\ead{vlefkopo@gmail.com},
\author[ETHZ_IfA,UBC]{Maryam Kamgarpour}\ead{mkamgar@ethz.ch}
\address[ETHZ_IfA]{Automatic Control Laboratory, ETH Zurich, Zürich 8092, Switzerland}
\address[UBC]{Electrical and Computer Engineering, University of British Columbia, Vancouver, V6Z 1Z4, Canada}

\begin{keyword}
	Autonomous vehicles; trajectory planning; stochastic optimal control; chance constraints.
\end{keyword}

\begin{abstract}
	We tackle the problem of trajectory planning in an environment comprised of a set of obstacles with uncertain time-varying locations. The uncertainties are modeled using widely accepted Gaussian distributions, resulting in a chance-constrained program. Contrary to previous approaches however, we do not assume perfect knowledge of the moments of the distribution, and instead estimate them through finite samples available from either sensors or past data. We derive tight concentration bounds on the error of these estimates  to sufficiently tighten the chance-constraint program. As such, we provide provable guarantees on satisfaction of the chance-constraints corresponding to the nominal yet unknown moments. We illustrate our results with two autonomous vehicle trajectory planning case studies.
\end{abstract}

\end{frontmatter}

\section{Introduction} \label{sec:introduction}
A major challenge in real-world employment of autonomous systems such as robots and autonomous cars is handling uncertainties in the environment.  For example, from the perspective of the autonomous car, the driving behavior of the nearby cars is  to a large extent unpredictable. In a robotic search-and-rescue, the location of goals and obstacles are a priori uncertain. In several applications of autonomous vehicles, large number of past data on the environment conditions, e.g. driving behavior of cars or environment maps exist either through camera or LIDAR measurements ~\cite{Kendall2017,Feng2018}. In either case, a large number of samples (e.g.\ potential positions of obstacles) is made available, leading to a requirement for an approach that reasons about the magnitude of the uncertainty based on these samples and ensures safety while allowing for real-time computation~\cite{Michelmore2019}. The approach proposed in this work addresses this issue of generating safe trajectories while working with an arbitrary finite number of samples. 

Past work in trajectory planning considers uncertainties stemming from two sources: 1) the autonomous agent's dynamic and measurement model~\cite{Ono2010,Blackmore2011,Vitus2012,Ono2015,Raman2015}; 2) the safe set in the environment ~\cite{Carvalho2014,Jha2018,Sessa2018}. While the former uncertainty is widely researched, the latter is more realistic in emerging applications such as autonomous driving. Trajectory planning in both formulations of uncertainty  can be addressed with robust optimization, ensuring constraint satisfaction for all possible uncertainty realization. This approach can  be too conservative. Chance-constrained formulation overcomes this conservativeness at the price of allowing (an arbitrary small) constraint violation probability. Since the support of the uncertainty sets is often unknown whereas sample data or probabilistic models of the uncertainty might be available, a chance-constrained formulation can better capture the problem data and offer a reasonable solution.

Depending on the assumptions on the uncertainty probability distribution, the chance-constrained program is handled often either with deterministic reformulations or the scenario approach. Blackmore et al.~\cite{Blackmore2011} consider polyhedral \emph{deterministic} obstacles and a linear plant model perturbed by Gaussian noise with \emph{known} moments. Accordingly, they reformulate the chance constraints and employ a specialized branch-and-bound technique to solve the resulting disjunctive program of obstacle avoidance. This approach is not directly extendable to stochastic obstacles as it would result in a nonlinear variant of the risk allocation problem. Jha et al.~\cite{Jha2018} consider polyhedral \emph{uncertain} obstacles whose stochasticity is normally distributed with \emph{known} moments and a \emph{deterministic} plant model that is to be controlled. They reformulate the chance constraints assuming a Gaussian distribution and then encode the obstacle avoidance problem as a mixed-integer scheme. The above methods work under the assumption of perfect knowledge of the Gaussian moments, whereas in reality one can only estimate an uncertainty model based on sensor measurements or past data.

Sessa et al.~\cite{Sessa2018} consider noise entering nonlinearly into the control system. They do not make assumptions about the noise distribution, but instead consider that a sufficient number of samples of it are available. As such, using the scenario approach, they provide probabilistic guarantees of the designed disturbance feedback control. While their work produces control policies that can adhere to safety under general disturbance models, a significant number of samples are required for the approach to produce any safety guarantees. Furthermore, the approach is too computationally intense to be implemented in real time.

If the objectives on the control system are safety or reachability, then a stochastic reachability problem can be formulated~\cite{Summers2013,Ono2015}. Summers et al.~\cite{Summers2013} consider a controller synthesis problem with a reach-avoid objective on stochastic sets. Ono et al.~\cite{Ono2015} consider a chance-constrained dynamic programming problem with general objective functions and safety constraints. The chance constraints are simplified using the Boole's inequality and the solution to the problem is approximated through a root-finding algorithm on its dual reformulation. The plant dynamics are nonlinear and perturbed by noise with a \emph{known} probability distribution.

We attempt to reach a compromise between the purely data-driven approach of~\cite{Sessa2018} and the known distribution assumptions of~\cite{Jha2018} for trajectory generation in stochastic environments. In particular, we take on the fairly accepted model of a Gaussian distributed uncertainty but consider the realistic case in which the moments of the distribution are \emph{unknown} and are estimated through finite samples. Such samples are often obtained through sensors, e.g. LIDAR or camera, or past collected data, e.g. historic driving trajectories. The choice of Gaussian distribution is on the one hand due to the fact that it lends itself to computationally tractable and analytic reformulations. On the other hand, it has several practical applications. Sensor noise, and more generally any unimodal distribution, is often approximated with a Gaussian distribution. Moreover, Gaussian noise is increasingly used in autonomous driving applications to model perturbations from nominal prototype maneuvers when predicting the behavior of other vehicles~\cite{Carvalho2014}.

While the problem of chance-constrained programming under uncertain moments has been addressed through the seminal work of~\cite{Calafiore2006a}, the case of tight concentration inequalities given a normally distributed model and incorporating this in a  stochastic trajectory planning were not addressed. To this end, we derive concentration bounds on the estimation error of the Gaussian moments,  propose a reformulation of the chance-constrained problem using only the moment estimates and  prove its solution's feasibility (in a probabilistic sense) for the original problem. We also extend the proposed method to consider \emph{dynamic} uncertain obstacles and implement it in closed loop through a receding horizon scheme. Early results of the work in this paper were presented in a conference paper in~\cite{Lefkopoulos2019}. Our work here completes the early studies by extending it in the following ways:
\begin{enumerate}
	\item Providing improved moment concentration bounds, as presented in Section~\ref{sec:momentConcentrationInequalities} and thus reducing the conservativeness  in~\cite{Lefkopoulos2019}.
	\item Extending the open-loop approach in ~\cite{Lefkopoulos2019} to a closed-loop control scheme, as presented in Section~\ref{sec:recedingHorizonImplementation}.
	\item Providing two new case studies to illustrate the above contributions.
\end{enumerate}

The rest of this paper is organized as follows. Section~\ref{sec:problemStatement} states the problem and reformulates it as a mixed integer second order cone program. Our result on incorporation of the moment uncertainties through concentration bounds is presented in Section~\ref{sec:momentsRobustApproach}. In Section~\ref{sec:recedingHorizonImplementation} a receding horizon scheme utilizing our approach is proposed. Section~\ref{sec:simulations} demonstrates the approach with two case studies in autonomous vehicles. Finally, we conclude in Section~\ref{sec:conclusion}.

\subsection*{Notation}
The default vector type is a column vector, denoted as $x \in \Reals^{n}$, with a row vector being denoted as $x \in \Reals^{1 \times n}$. We denote the set of positive natural numbers by $\Nats_{>0}$ and the set of positive real numbers by $\Reals_{>0}$. We denote the subset from $a$ to $b$ of an ordered set $\mathbb{X}$ by $\mathbb{X}[a,b]$, with $[a,b]$ being equivalent to $\Reals[a,b]$. We denote a conjunction (logical AND operator) by $\wedge$ and a disjunction (logical OR operator) by $\vee$. Given two row vectors $a \in \Reals^{1 \times n}$ and $b \in \Reals^{1 \times m}$, we denote their concatenation (which is also a row vector) as $[a,b] \in \Reals^{1 \times (n+m)}$. By $\Gauss(\mu,\Sigma)$ we denote the $n$-dimensional multivariate Gaussian distribution with mean $\mu \in \Reals^n$ and covariance $\Sigma \in \Reals^{n \times n}$. By $\Psi^{-1}(\cdot)$ we denote the inverse cumulative distribution function of $\Gauss(0,1)$. We denote a symmetric positive definite matrix by $A \succ 0$. A set of random variables $d_i$ are i.i.d.\ if they are independent and identically distributed. $I_n \in \Reals^{n \times n}$  denotes the $n$\nobreakdash-dimensional identity matrix. By $\mymathbb{0}$ we denote a zero vector or matrix of appropriate dimensions.

\section{Problem Statement} \label{sec:problemStatement}
We consider the system evolving according to the  time-varying and linear dynamics:
\begin{equation} \label{eq:systemDynamicsLTV}
	x_{t+1} = A_t x_t + B_t u_t \,,
\end{equation}
where $x_t \in \Reals^{n_x}$ is the state, $u_t \in \Reals^{n_u}$ is the input and $A_t \in \Reals^{n_x \times n_x}$, $B_t \in \Reals^{n_x \times n_u}$ are the system dynamics' matrices at time $t \in \Nats$. The time-varying model captures linearization of a nonlinear dynamics around a nominal trajectory and thus, significantly generalizes the applicability of the linear model. Given a horizon length $N \in \Nats_{>0}$, we denote the finite-length input sequence as $\bm{u} \coloneqq (u_0,\dots,u_{N-1}) \in \Reals^{N n_u}$. Given an initial state $x_0$ and $\bm{u}$, the evolution of the state trajectory is denoted as $\bm{x} \coloneqq (x_1,\dots,x_N) \in \Reals^{N n_x}$. The  control inputs are constrained to a convex set $\mathcal{U}$, $u_t \in \mathcal{U}$ for all $t \in  \Nats[0,N-1]$. The assumption of the set $\mathcal{U}$ being time-invariant is only for simplicity in notation.

For trajectory planning in uncertain environments, the state needs to avoid a set of obstacles with \emph{uncertain} locations. We consider obstacles modeled by  polyhedrons. Let $N_o \in \Nats$ be the number of obstacles (indexed by $j$), with $F_j$ number of faces (indexed by $i$). The $j$\nobreakdash-th obstacle's interior $\mathcal{O}_j^t \in \Reals^{n_x}$ at time $t$ can be expressed as the conjunction of the linear constraints of its faces:
\begin{equation} \label{eq:obstacleRegion}
	\mathcal{O}^t_j \coloneqq \left\{ x \in \Reals^{n_x} : \wedge_{i=1}^{F_j}  {a^t_{ij}}^\Trans x + b^t_{ij} \leq 0 \right\} \,,
\end{equation}
where $a^t_{ij} \in \Reals^{n_x}$ and $b^t_{ij} \in \Reals$ define the $j$\nobreakdash-th's obstacle $i$\nobreakdash-th face at time $t$, and are concatenated in $d^t_{ij} \coloneqq [ {a^t_{ij}}^\Trans, b^t_{ij} ]^\Trans \in \Reals^{n_x+1}$. The complement of the obstacle set is defined via the disjunction of the constraints:
\begin{equation} \label{eq:obstacleRegionComplement}
	\overline{\mathcal{O}^t_j} \coloneqq \left\{ x \in \Reals^{n_x} : \vee_{i=1}^{F_j}  {a^t_{ij}}^\Trans x + b^t_{ij} > 0 \right\} \,.
\end{equation}
The safe set over the planning horizon is $\bm{\mathcal{X}} \subset \Reals^{N n_x}$:
\begin{align} \label{eq:stateSafeSet}
	\bm{\mathcal{X}} & \coloneqq \left\{ \bm{x} \in \Reals^{N n_x} : \wedge_{t=1}^{N} x_t \in \mathcal{X}_t \right\}\,,\\
	\label{eq:stateSafeSetTime}
	\mathcal{X}_t & \coloneqq \left\{ x \in \Reals^{n_x} : \wedge_{j=1}^{N_o} \vee_{i=1}^{F_j} {a^t_{ij}}^\Trans x + b^t_{ij} > 0 \right\}\,.
\end{align}

The uncertainty of the obstacles' locations is captured by considering  $d^t_{ij} \sim \mathcal{D}^t_{ij}$, where $\mathcal{D}^t_{ij}$ is a probability distribution on $\Reals^{n_x + 1}$. Thus, the set $\bm{\mathcal{X}}$ is stochastic. Hence, as motivated in the introduction we formulate the safety requirement as a \emph{joint chance constraint}:
\begin{equation} \label{eq:stateSafeSetConstraint}
	\Pr(\bm{x} \in \bm{\mathcal{X}}) \geq 1-\epsilon \,,
\end{equation}
where $\epsilon$ is a prescribed safety margin. 

We assume the objective of the autonomous vehicle is captured through a convex cost function $J(x_0,\cdot): \Reals^{n_u} \rightarrow \Reals$. Thus, the chance-constraint safe trajectory planning is:
\begin{mini!}<b>
	{\bm{u}}{J(x_0,\bm{u}) \label{eq:initialProblemCost}}
	{\label{eq:initialProblem}}{}
	\addConstraint{\bm{x},\bm{u} \text{ satisfy~\eqref{eq:systemDynamicsLTV} with initial state } x_0}{\label{eq:initialProblemDynamics}}
	\addConstraint{\bm{u} \in \bm{\mathcal{U}}} {\label{eq:initialProblemInputConstraints}}
	\addConstraint{\Pr(\bm{x} \in \bm{\mathcal{X}}) \geq 1-\epsilon}{\label{eq:initialProblemChanceConstraint}}
\end{mini!}
where $\bm{\mathcal{U}} \coloneqq \mathcal{U} \cross \dots \cross \mathcal{U} \subseteq \Reals^{N n_u}$. Problem~\eqref{eq:initialProblem} is not in a form that can be directly handled by off\nobreakdash-the-shelf optimization solvers because of the chance constraint~\eqref{eq:initialProblemChanceConstraint}. As such, below we recall the approach to reformulate this constraint into a tractable deterministic form that depends on the moments of the distributions $\mathcal{D}^t_{ij}$.

\subsection{Single chance constraints formulation} \label{sec:singleChanceConstraintsFormulation}

\textbf{Big-M method.} \label{sec:bigM}
A common technique to deal with the non-convexity of a disjunction such as in~\eqref{eq:obstacleRegionComplement} is the Big-M reformulation~\cite{Schouwenaars2001,Richards2002}. By introducing one binary variable per obstacle face, satisfaction of~\eqref{eq:obstacleRegionComplement} is equivalent to:
\begin{equation} \label{eq:constraintBigM}
    x_t \in \overline{\mathcal{O}^t_j} \Leftrightarrow
\end{equation}
where $z^t_{ij} \in \Bins$ and $M \in \Reals_{>0}$,  with $M$ chosen as a very large constant so that 
$M > \sup_{x_t} -{a^t_{ij}}^\Trans x_t - b^t_{ij}$. By this choice, $z^t_{ij} = 1$ implies that the $i$-th constraint is  vacuous since it holds regardless of value of the decision variable $x_t$, whereas $z^t_{ij} = 0$ for at least one $i$ ensures~\eqref{eq:obstacleRegionComplement}.  Using~\eqref{eq:constraintBigM} we equivalently write constraint~\eqref{eq:initialProblemChanceConstraint} as
\begin{align} \label{eq:chanceConstraintBigM}
	&\Pr(\wedge_{t=1}^{N} \wedge_{j=1}^{N_o} \wedge_{i=1}^{F_j} {a^t_{ij}}^\Trans x_t + b^t_{ij} + M z^t_{ij} > 0) \geq 1-\epsilon\, , \\
		\label{eq:constraintBinaryVariables}
	&\sum_{i=1}^{F_j} z^t_{ij} < F_j,\quad \forall t \in \Nats[1,N],\quad \forall j \in \Nats[1,N_o].
\end{align}

\textbf{Boole's inequality.} \label{sec:boolesInequality}
Constraint~\eqref{eq:chanceConstraintBigM} is a joint chance constraint due to the requirement of probability of a joint set of events satisfying a threshold. Joint chance constraints are hard to deal with~\cite{Nemirovski2007}. Simply evaluating a joint chance constraint, given $\bm{x}$, requires calculating a multivariate integral which in general can only be performed for low dimensions. Furthermore, generally there is no tractable way to approximate them. 
Based on  Boole's inequality~\cite{Casella2002}:
\begin{equation} \label{eq:boolesInequality}
	\Pr(\vee A_i) \leq \sum_i \Pr(A_i) \,.
\end{equation}
for events $A_i$, one can split up a joint chance constraint into multiple single chance constraints. The single chance-constraints are easier to analyze at the price of conservativeness introduced in the inequality above.  

Using~\eqref{eq:boolesInequality} in complements form, a sufficient condition for satisfying constraint~\eqref{eq:chanceConstraintBigM} is~\cite{Jha2018}:
\begin{equation} \label{eq:singleChanceConstraints}
	\wedge_{t=1}^{N} \wedge_{j=1}^{N_o} \wedge_{i=1}^{F_j} \Pr({a^t_{ij}}^\Trans x_t + b^t_{ij} + M z^t_{ij} > 0) \geq 1-\epsilon^t_{ij} \,,
\end{equation}
where $\epsilon^t_{ij} \in (0,0.5)$ are risk variables that must satisfy
\begin{equation} \label{eq:riskAllocationConstraint}
	\sum_{t=1}^{N} \sum_{j=1}^{N_o} \sum_{i=1}^{F_j} \epsilon^t_{ij} \leq \epsilon \,.
\end{equation}

\subsection{Risk allocation} \label{sec:riskAllocation}
Risk allocation addresses choosing the probability of violation for individual chance constraints, $\epsilon^t_{ij}$ such that ~\eqref{eq:riskAllocationConstraint} is satisfied. A simple although generally conservative way to allocate the risk is to do so uniformly~\cite{Blackmore2006,Nemirovski2007}:
\begin{equation} \label{eq:uniformRiskAllocation}
\epsilon_{\text{uni}}:= \epsilon^t_{ij} = \frac{\epsilon}{N \sum_{j=1}^{N_o} F_j}.
\end{equation}
In recent literature, different approaches for improving the risk allocation have been investigated. In~\cite{Vitus2011} an iterative procedure is proposed, solving the optimization problem using a fixed risk allocation and then optimizing over the risk allocation itself. In~\cite{Jha2018} a heuristics approach is designed, by transferring cost from the vacuous constraints to the non\nobreakdash-vacuous ones. Optimizing simultaneously over both the trajectory generation and the risk allocation has also been investigated~\cite{Blackmore2009,Ono2010,Blackmore2011}. The latter approaches however result in a non-convex optimization problem due to multiplication of the original decision variables with the risk variables. 

Our first result below shows that, for the problem at hand, one can significantly improve upon the uniform risk allocation without the need for additional optimization. 

\begin{lem} \label{lem:improvedRiskAllocation}
	Using the risk allocation:
	\begin{equation} \label{eq:improvedRiskAllocation}
		\epsilon^t_{ij} = \frac{\epsilon}{N N_o} > \epsilon_{\text{uni}}\,,
	\end{equation}
	and the binary variable constraints
\begin{equation} \label{eq:improvedRiskProblemBinaryConstraints}
\sum_{i=1}^{F_j} z^t_{ij} = F_j-1 \,,
\end{equation}

any feasible point for problem $\mathcal{P}\coloneqq\{\eqref{eq:singleChanceConstraints},~\eqref{eq:improvedRiskAllocation},~\eqref{eq:improvedRiskProblemBinaryConstraints}\}$ is also feasible for problem $\mathcal{P}'\coloneqq\{\eqref{eq:constraintBinaryVariables},~\eqref{eq:singleChanceConstraints},~\eqref{eq:riskAllocationConstraint}\}$ and vice versa. 
\end{lem}

\begin{pf}
	First, let us show that any feasible point for $\mathcal{P}$ is feasible for $\mathcal{P'}$. Notice that every binary variable $z^t_{ij}$ that satisfies~\eqref{eq:improvedRiskProblemBinaryConstraints} also satisfies~\eqref{eq:constraintBinaryVariables}. For a given obstacle $j$, let $i^*$ denote the unique binary term above such that $z^t_{i^*j} = 0$. It follows that for all $i \neq i^*$, the corresponding constraints $a^t_{ij} x_t + b^t_{ij} + Mz^t_{ij}$ holds regardless of the values  the random variables take due to the choice of $M$ in the big-$M$ method. Consequently, \eqref{eq:improvedRiskAllocation} is sufficient to ensure~\eqref{eq:singleChanceConstraints}. It follows that any feasible point for $\mathcal{P}$ is also feasible for $\mathcal{P'}$. 
	
	Next, let us show that any feasible point for $\mathcal{P'}$ is feasible for $\mathcal{P}$. Suppose there exists an index $i$ such that ${z'}^t_{ij} = 0$ in problem $\mathcal{P'}$ with its corresponding $x'_t$,  whereas this term is  $z^t_{ij} = 1$ in $\mathcal{P}$. From the choice of $M$ in big-$M$ method, it follows that $x'_t$ also satisfies constraint $i$ in $\mathcal{P}$ for any values of the random variables (since $z^t_{ij} =1$  means that $x_t$ can be set arbitrarily.) Hence, the risk corresponding to these constraints can be set to zero. 
\end{pf}

\subsection{Mixed-integer second-order cone formulation} \label{sec:chanceConstraintReformulation}
Consistent with several existing trajectory planning works~\cite{Blackmore2011,Carvalho2014,Jha2018}, we assume Gaussian uncertainties to allow the analytic reformulation of the chance constraints while capturing general uncertainties with a unimodal and unbounded support set.

\begin{assum} \label{ass:gaussianDistribution}
	Each $d^t_{ij}$ is a Gaussian random variable, $d^t_{ij} \sim \mathcal{N}(\mu^t_{ij},\Sigma^t_{ij})$.
\end{assum}

Under Assumption~\ref{ass:gaussianDistribution}, each chance constraint of~\eqref{eq:singleChanceConstraints} is equivalent to the following second-order cone constraint~\cite[Theorem~2.1]{Calafiore2006a}:
\begin{equation} \label{eq:reformulatedChanceConstraint}
	\Psi^{-1}(1-\epsilon^t_{ij}) \norm{(\Sigma^t_{ij})^{\flatfrac{1}{2}} \tilde{x}_t}_2 \leq {\mu^t_{ij}}^\Trans \tilde{x}_t + M z^t_{ij} \,,
\end{equation}
where $\tilde{x} \coloneqq [x^\Trans, 1]^\Trans \in \Reals^{n_x+1}$.

Finally, as a direct result of Sections~\ref{sec:singleChanceConstraintsFormulation},~\ref{sec:riskAllocation} and~\ref{sec:chanceConstraintReformulation}, we can conservatively satisfy the joint chance constraint~\eqref{eq:initialProblemChanceConstraint} through the implications:
\begin{equation} \label{eq:reformulationEquivalence}
	\eqref{eq:initialProblemChanceConstraint} \Leftrightarrow \eqref{eq:chanceConstraintBigM} \wedge \eqref{eq:constraintBinaryVariables} \Leftarrow \eqref{eq:reformulatedChanceConstraint} \wedge \eqref{eq:improvedRiskProblemBinaryConstraints}
\end{equation}

\section{Moments Robust Approach} \label{sec:momentsRobustApproach}
To ensure satisfaction of the chance constraints~\eqref{eq:singleChanceConstraints} when the moments $\mu^t_{ij}$ and $\Sigma^t_{ij}$ are estimated from data, the estimates' uncertainties need to be accounted for in the constraint formulation. Let us start with a simple example showing that if the uncertainty in the moments estimation  is not incorporated, we obtain a lower bound on the probability of constraint violation.  
\begin{exmp} \label{ex:MRA}
	Consider the chance-constrained problem:
	\begin{mini!}<b>
		{x \in \Reals}{x}
		{\label{eq:exampleProblem}}{}
		\addConstraint{\Pr(x \geq \delta) \geq 1-\epsilon}{\label{eq:exampleProblemConstraint}}
	\end{mini!}
	where $\delta \sim \Gauss(\mu,\sigma^2)$. As established in Section~\ref{sec:chanceConstraintReformulation}, the chance constraint~\eqref{eq:exampleProblemConstraint} is equivalent to the constraint:
	\begin{equation} \label{eq:exampleDeterministicConstraint}
		x \geq \mu + \Psi^{-1}(1-\epsilon) \sigma \,,
	\end{equation}
	hence making the solution of Problem~\eqref{eq:exampleProblem} apparent:
	\begin{equation} \label{eq:exampleProblemSolution}
		x^* = \mu + \Psi^{-1}(1-\epsilon) \sigma \,.
	\end{equation}
	Often in practice, the mean value $\mu$ is unknown and is estimated from $N_s$ samples using the estimate of~\eqref{eq:sampleMeanEstimate}. This leads to the \emph{approximate} solution:
	\begin{equation} \label{eq:exampleApproximateSolution}
		\hat{x}^* = \hat{\mu} + \Psi^{-1}(1-\epsilon) \sigma \,.
	\end{equation}
	Since $\hat{\mu}$ is a random variable, the solution~\eqref{eq:exampleApproximateSolution} is itself also a random variable. The probability that $\hat{x}^*$ violates the original constraint~\eqref{eq:exampleProblemConstraint}, denoted by $\delta_{\text{viol}}$, is:
	\begin{equation*}
		\eta_{\text{viol}} = \Pr(\hat{\mu} + \Psi^{-1}(1-\epsilon) \sigma < \mu + \Psi^{-1}(1-\epsilon) \sigma) = \Pr(\hat{\mu} < \mu) \,.
	\end{equation*}
	 It is known that the sample mean $\hat{\mu}$ is distributed as:
	\begin{equation} \label{eq:exampleMeanConcentration}
		\frac{\hat{\mu}-\mu}{\sigma^2} \sqrt{N_s} \sim t_{N_s-1}
	\end{equation}
	where $t_k$ denotes the Student's t-distribution with $k$ degrees of freedom. The t-distribution is symmetric, which leads us to the conclusion that $\eta_{\text{viol}} = 0.5$. Thus, if we solve Problem~\eqref{eq:exampleProblem} by estimating the mean value of $\delta$ from samples, \emph{regardless of the number of samples}, there is a \SI{50}{\percent} chance that the original constraint is violated. 
\end{exmp}

To incorporate the uncertainty in the moments' estimates and guarantee constraint satisfaction with high confidence, in Section~\ref{sec:momentConcentrationInequalities} we derive tight moment concentration bounds. In Section~\ref{sec:robustifyingChanceConstraints} we reformulate  the original optimization problem based on the above. For the sake of brevity the indices $t$, $i$, and $j$ are omitted from the random variable $d^t_{ij}$ and its moments. 

\subsection{Moment concentration inequalities} \label{sec:momentConcentrationInequalities}
Let us consider having access to $N_s$ i.i.d.\ samples $d_1,\dots,d_{N_s}$ of $d$. We form the sample mean and covariance estimates:
\begin{subequations} \label{eq:sampleEstimates}
\begin{align}
	\hat{\mu} & = \frac{1}{N_s} \sum_{i=1}^{N_s} d_i \,, \label{eq:sampleMeanEstimate} \\
	\hat{\Sigma} & = \frac{1}{N_s-1} \sum_{i=1}^{N_s} (d_i-\hat{\mu})(d_i-\hat{\mu})^\Trans \,. \label{eq:sampleCovarianceEstimate}
\end{align}
\end{subequations}

\begin{assum} \label{ass:positiveDefiniteSampleCovariance}
	$\hat{\Sigma}$ is positive definite, i.e. $\hat{\Sigma} \succ 0$\footnote{Assumption~\ref{ass:positiveDefiniteSampleCovariance} is not reasonable for a very small number of samples $N_s < n$, since the sample covariance matrix is a sum of $N_s$ rank-1 matrices. In the the trajectory generation application, however, it is reasonable to assume that $N_s$ is sufficiently larger than $n$ as will be discussed in the examples.}.
\end{assum}

The concentration bound $r_1$ of Lemma~\ref{lem:sampleMeanBound} corresponding to the $\mu$ is identical to the one derived in~\cite[Lemma~1]{Lefkopoulos2019}, and hence its proof is omitted. The new concentration bound $r_2$ of Lemma~\ref{lem:covarianceBound} corresponding to $\Sigma$ improves the one in ~\cite[Lemma~2]{Lefkopoulos2019} (cf.\ Remark~\ref{rem:concentrationBoundAsymptotic}).

\begin{lem}[from \cite{Lefkopoulos2019}] \label{lem:sampleMeanBound}
	Under Assumptions~\ref{ass:gaussianDistribution}--\ref{ass:positiveDefiniteSampleCovariance} and using the sample estimates~\eqref{eq:sampleMeanEstimate} and~\eqref{eq:sampleCovarianceEstimate},  with probability $1-\beta$, $\beta \in (0,1)$:
	\begin{equation} \label{eq:sampleMeanBound}
		\norm{\mu-\hat{\mu}}_2 \leq r_1 \coloneqq \sqrt{\frac{T_{n,N_s-1}^2(1-\beta)}{N_s \lambda_{\min}(\hat{\Sigma}^{-1})}} \,,
	\end{equation}
	where $T^2_{a,b}(p)$ denotes the $p$\nobreakdash-th quantile of the Hotelling's T\nobreakdash-squared distribution~\cite{Hotelling1931} with parameters $a$ and $b$.
\end{lem}

\begin{lem} \label{lem:covarianceBound}
	Under Assumptions~\ref{ass:gaussianDistribution}--\ref{ass:positiveDefiniteSampleCovariance} and using the sample estimates~\eqref{eq:sampleMeanEstimate} and~\eqref{eq:sampleCovarianceEstimate}, define the constant:
	\begin{equation} \label{eq:sampleCovarianceBoundDefinition}
	\begin{split}
		r_{2} \coloneqq \max\Big\{ 
		& \abs\big{ 1-\frac{N_s-1}{\chi^2_{N_s-1,1-\flatfrac{\beta}{2}}} } , \abs\big{ 1-\frac{N_s-1}{\chi^2_{N_s-1,\flatfrac{\beta}{2}}} } \Big\} \,,
	\end{split}
	\end{equation}
	where $\chi^2_{k,p}$ is the $p$\nobreakdash-th quantile of the chi-squared distribution with $k$ degrees of freedom. Then, with probability $1-\beta$, $\beta \in (0,1)$:
	\begin{equation} \label{eq:sampleCovarianceBound}
		\abs{x^\Trans (\Sigma - \hat{\Sigma}) x} \leq x^\Trans \hat{\Sigma} x r_2.
	\end{equation}
\end{lem}

\begin{pf}
	It is known~\cite{Mardia1979} that $(N_s-1)\hat{\Sigma}$ follows a Wishart distribution, i.e.:
	\begin{equation} \label{eq:sampleCovarianceDistribution}
		(N_s-1)\hat{\Sigma} \sim W_n(\Sigma,N_s-1) \,,
	\end{equation}
	where $W_n(\Sigma,N_s-1)$ denotes the Wishart distribution with positive definite scale matrix $\Sigma$ and $N_s-1$ degrees of freedom associated with an $n$\nobreakdash-variate normal distribution. If  $A \in \Reals^{n \times n}$ follows a Wishart distribution $W_n(\Sigma,m)$ and $x \in \Reals^{n}$ is a nonzero vector, then~\cite[p.~535]{Rao1965}:
	\begin{equation} \label{eq:quadraticWishartDistribution}
		x^\Trans A x \sim x^\Trans \Sigma x \chi^2_m \,,
	\end{equation}
	where $\chi^2_m$ is the chi-squared distribution with $m$ degrees of freedom. Combining~\eqref{eq:sampleCovarianceDistribution} and~\eqref{eq:quadraticWishartDistribution} we obtain:
	\begin{equation} \label{eq:quadraticCovarianceDistribution}
		x^\Trans \hat{\Sigma} x \sim \frac{1}{N_s-1} x^\Trans \Sigma x \chi^2_{N_s-1} \,.
	\end{equation}
	Hence, in a similar fashion to~\cite[p.~133]{Krishnamoorthy2006}, we can construct the $1-\beta$ confidence interval for~\eqref{eq:quadraticCovarianceDistribution}:
	\begin{equation} \label{eq:quadraticCovarianceInterval}
		x^\Trans \Sigma x \in \left[ \frac{(N_s-1)x^\Trans \hat{\Sigma} x}{\chi^2_{N_s-1,1-\flatfrac{\beta}{2}}}, \frac{(N_s-1)x^\Trans \hat{\Sigma} x}{\chi^2_{N_s-1,\flatfrac{\beta}{2}}} \right] \,,
	\end{equation}
	holds with probability $1-\beta$. By subtracting $x^\Trans \hat{\Sigma} x$ from~\eqref{eq:quadraticCovarianceInterval} it follows that:
	\begin{align*}
		x^\Trans (\Sigma - \hat{\Sigma}) x & \in \left[ \frac{(N_s-1)x^\Trans \hat{\Sigma} x}{\chi^2_{N_s-1,1-\flatfrac{\beta}{2}}} - x^\Trans \hat{\Sigma} x, \frac{(N_s-1)x^\Trans \hat{\Sigma} x}{\chi^2_{N_s-1,\flatfrac{\beta}{2}}} - x^\Trans \hat{\Sigma} x \right] \\
		& = \left[ x^\Trans \hat{\Sigma} x (\frac{(N_s-1)}{\chi^2_{N_s-1,1-\flatfrac{\beta}{2}}} - 1), x^\Trans \hat{\Sigma} x (\frac{(N_s-1)}{\chi^2_{N_s-1,\flatfrac{\beta}{2}}} - 1) \right] \\
		& = x^\Trans \hat{\Sigma} x \left[ \frac{(N_s-1)}{\chi^2_{N_s-1,1-\flatfrac{\beta}{2}}} - 1, \frac{(N_s-1)}{\chi^2_{N_s-1,\flatfrac{\beta}{2}}} - 1 \right] \,,
	\end{align*}
	from which the statement of the lemma readily follows.
\end{pf}

\begin{figure}[t!]
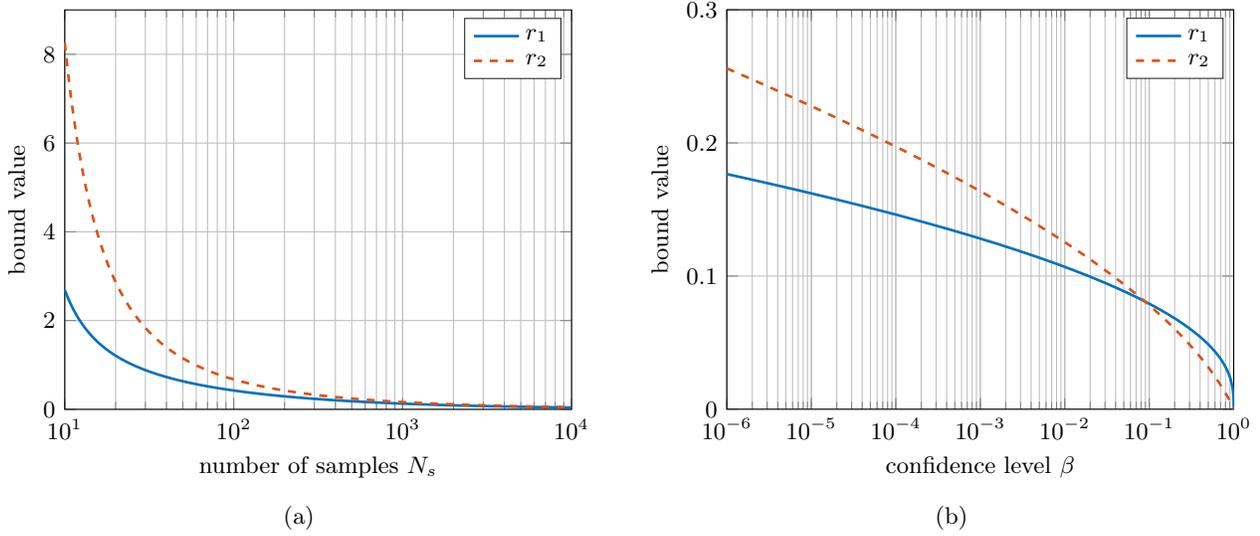

	\centering
	\setlength{\figureheight}{0.3\columnwidth}
	\setlength{\figurewidth}{0.4\columnwidth}
	\subfloat[]{
		\includegraphics{Example_sampleDependence.tikz}
		\label{fig:Example_sampleDependence}
	}
	\hfil
	\subfloat[]{
		\includegraphics{Example_betaDependence.tikz}
		\label{fig:Example_betaDependence}
	}
	\caption{(a): Concentration bounds $r_1$ (blue) and $r_2$ (red) for (a) a varying number of samples $N_s$ (logarithmic axis) and (b) a varying confidence $\beta$ (logarithmic axis). (a): The numerical values $n=3$ and $\beta = \num{e-3}$ were set to calculate the bounds. (b): The numerical values $n=3$, $N_s=\num{e3}$ and $\hat{\Sigma}=I_3$ were used to calculate the bounds.}
\end{figure}

\begin{rem} \label{rem:concentrationBoundAsymptotic}
	The concentration bounds $r_1$ and $r_2$ of~\eqref{eq:sampleMeanBound} and~\eqref{eq:sampleCovarianceBoundDefinition} asymptotically converge to zero as the number of samples $N_s$ grows. The convergence rate is illustrated in Fig.~\ref{fig:Example_sampleDependence}. In contrast, the concentration bounds in previous works, including the one provided in ~\cite[Lemma~2]{Lefkopoulos2019}, is less tight and does not asymptotically converge to zero.
\end{rem}

Going back to Example~\ref{ex:MRA}, the proposed Moments Robust Approach will lead to the solution:
	\begin{equation} \label{eq:exampleProblemSolutionMRA}
		\hat{x}^* = \hat{\mu} + \Psi^{-1}(1-\epsilon) \sigma + t_{N_s-1,1-\flatfrac{\beta}{2}} \frac{\sigma}{\sqrt{N_s}} \abs{\hat{\mu}} \,,
	\end{equation}
	where $t_{k,p}$ is the $p$\nobreakdash-th quantile of the  Student's t-distribution with $k$ degrees of freedom. Solution~\eqref{eq:exampleProblemSolutionMRA} is feasible with probability   $\eta_{\text{viol}} = \beta$, where $\beta$ trades off optimality of the $\hat{x}^*$ and the constraint violation.

\subsection{Robustifying the chance constraints} \label{sec:robustifyingChanceConstraints}
Using  Lemmas~\ref{lem:sampleMeanBound} and~\ref{lem:covarianceBound} on the concentration of the sample moments around their true values, we conservatively reformulate a chance constraint as follows.

\begin{lem} \label{lem:chanceConstraintReformulation}
	Under Assumptions~\ref{ass:gaussianDistribution}--\ref{ass:positiveDefiniteSampleCovariance} and using the sample estimates~\eqref{eq:sampleMeanEstimate} and~\eqref{eq:sampleCovarianceEstimate}, the chance constraint:
	\begin{equation} \label{eq:chanceConstraintExample}
		\Pr(d^\Trans \tilde{x} + Mz > 0) \geq 1-\epsilon \,,
	\end{equation}
	holds with a probability of at least $1-2\beta$, provided that:
	\begin{equation} \label{eq:mraReformulatedChanceConstraint}
		\Psi^{-1}(1-\epsilon) \sqrt{1+r_2} \norm{\hat{\Sigma}^{\flatfrac{1}{2}} \tilde{x}}_2 +r_1 \norm{\tilde{x}}_2 \leq {\hat{\mu}}^{\Trans} \tilde{x} + Mz \,.
	\end{equation}
\end{lem}

\begin{pf}
	In Section~\ref{sec:chanceConstraintReformulation} we established that the chance constraint~\eqref{eq:chanceConstraintExample} is equivalent to:
	\begin{equation} \label{eq:reformulatedChanceConstraintExample}
		\Psi^{-1}(1-\epsilon) \norm{\Sigma^{\flatfrac{1}{2}} \tilde{x}}_2 - {\mu}^\Trans \tilde{x} - Mz \leq 0 \,.
	\end{equation}
	According to Lemmas~\ref{lem:sampleMeanBound} and~\ref{lem:covarianceBound} the bounds~\eqref{eq:sampleMeanBound} and~\eqref{eq:sampleCovarianceBound} each hold with probability $1-\beta$ and thus hold jointly\footnote{By Cochran's theorem, for normal distributions the sample mean $\hat{\mu}$ and the sample covariance $\hat{\Sigma}$ are independent~\cite{Krishnamoorthy2006}.} with probability $(1-\beta)^2$. Thus, with probability $(1-\beta)^2 > 1-2\beta$:
	\begin{align*}
		\Psi^{-1}(1-\epsilon) \norm*{\Sigma^{\flatfrac{1}{2}} \tilde{x}}_2 - {\mu}^\Trans \tilde{x}_t - Mz {}={} & \Psi^{-1}(1-\epsilon) \sqrt{ {\tilde{x}}^\Trans \hat{\Sigma} \tilde{x} + {\tilde{x}}^\Trans (\Sigma-\hat{\Sigma}) \tilde{x} } - (\mu-\hat{\mu})^\Trans \tilde{x} - {\hat{\mu}}^\Trans \tilde{x} - Mz \\
		{}\leq{} & \Psi^{-1}(1-\epsilon) \sqrt{ {\tilde{x}}^\Trans \hat{\Sigma} \tilde{x} + {\tilde{x}}^\Trans \hat{\Sigma} \tilde{x} r_2 } + \norm*{\mu-\hat{\mu}}_2 \norm*{\tilde{x}}_2 - {\hat{\mu}}^\Trans \tilde{x} - Mz \\
		{}\leq{} & \Psi^{-1}(1-\epsilon) \sqrt{ {\tilde{x}}^\Trans \hat{\Sigma} \tilde{x} (1+r_2) } + r_1 \norm*{\tilde{x}}_2 - {\hat{\mu}}^\Trans \tilde{x} - Mz \\
		{}={} & \Psi^{-1}(1-\epsilon) \sqrt{1+r_2} \norm{\hat{\Sigma}^{\flatfrac{1}{2}} \tilde{x}}_2 + r_1 \norm*{\tilde{x}}_2 - {\hat{\mu}}^\Trans \tilde{x} - Mz \,,
	\end{align*}
	where the first inequality is an application of the Cauchy-Schwarz inequality and the second inequality follows from Lemmas~\ref{lem:sampleMeanBound} and~\ref{lem:covarianceBound}.
\end{pf}

We are now ready to address Problem~\eqref{eq:initialProblem} given sample data on the position of obstacles.
\begin{mini!}<b>
	{\bm{u},z^t_{ij}}{J(x_0,\bm{u}) \label{eq:mraProblemCost}}
	{\label{eq:mraProblem}}{}
	\addConstraint{\bm{x},\bm{u} \text{ satisfy~\eqref{eq:systemDynamicsLTV} with initial state } x_0}{\label{eq:mraDynamics}}
	\addConstraint{\bm{u} \in \bm{\mathcal{U}}} {\label{eq:mraProblemInputConstraints}}
	\addConstraint{\Psi^{-1}(1-\epsilon^t_{ij}) \sqrt{1+r_2} \norm{({\hat{\Sigma}}^t_{ij})^{\flatfrac{1}{2}} \tilde{x}_t}_2 + r^t_{1,ij} \norm*{\tilde{x}_t}_2 \leq {\hat{\mu}_{ij}}^{t\Trans} \tilde{x}_t + Mz^t_{ij}}{\label{eq:mraProblemConstraints}}
	\addConstraint{\sum_{i=1}^{F_j} z^t_{ij} = F_j - 1}{\label{eq:mraProblemBinaryConstraints}}
\end{mini!}
where constraint~\eqref{eq:mraProblemConstraints} must hold for all $t \in \Nats[1,N], j \in \Nats[1,N_o], i \in \Nats[1,F_j]$ and constraint~\eqref{eq:mraProblemBinaryConstraints} must hold for all $t \in \Nats[1,N], j \in \Nats[1,N_o]$.

\begin{thm} \label{thm:mraProblemFeasibility}
	Under Assumptions~\ref{ass:gaussianDistribution}--\ref{ass:positiveDefiniteSampleCovariance}, using the sample estimates~\eqref{eq:sampleMeanEstimate} and~\eqref{eq:sampleCovarianceEstimate}, and the risk allocation~\eqref{eq:improvedRiskAllocation}, a solution to Problem~\eqref{eq:mraProblem} is a feasible solution to Problem~\eqref{eq:initialProblem} with a probability of at least $ 1 - 2 \beta N N_o$. Furthermore, the solution of Problem~\eqref{eq:mraProblem} asymptotically converges to the solution of the exact moment problem, with constraint~\eqref{eq:reformulatedChanceConstraint} instead of~\eqref{eq:mraProblemConstraints}, as the number of available samples $N_s$ converges to infinity. 
\end{thm}

\begin{pf}
	According to Lemma~\ref{lem:chanceConstraintReformulation} each constraint of~\eqref{eq:mraProblemConstraints} implies the corresponding constraint~\eqref{eq:singleChanceConstraints} with probability $1-2\beta$. By requiring that this implication holds jointly for all $k = N N_o$ non-vacuous constraints, and noting that $(1-2\beta)^k > 1 - 2 \beta k$,\footnote{By considering the function $f(\beta)=(1-2\beta)^k+2k\beta -1$ and verifying that $f(0)=0$ and $f'(\beta)>0$ for $k \geq 1$ and $0<\beta<\flatfrac{1}{2}$.} we can conclude that a solution to Problem~\eqref{eq:mraProblem} is a feasible solution to Problem~\eqref{eq:initialProblem} with a probability of at least $ 1 - 2 \beta N N_o$. We do not consider the vacuous constraints by applying the same reasoning as the proof of Lemma~\ref{lem:improvedRiskAllocation} but for $\beta$ instead of $\epsilon^t_{ij}$. The asymptotic convergence of solution to the case of exact moment knowledge follows Remark~\ref{rem:concentrationBoundAsymptotic} and is due to the improved bounds compared to~\cite{Lefkopoulos2019}.
\end{pf}

\begin{rem} \label{rem:concentrationBounds}
	The total confidence level $ 1 - 2 \beta N N_o $ of Theorem~\ref{thm:mraProblemFeasibility} can be set at a desired level by prescribing a sufficiently high confidence $1-\beta$. For example, for a total confidence of $\num{e-3}$ if $N=10$ and $N_o=5$ we would prescribe $\beta = \num{e-5}$. This is not prohibitive since the concentration bounds $r_1$ and $r_2$ of~\eqref{eq:sampleMeanBound} and~\eqref{eq:sampleCovarianceBoundDefinition} have favorable (approximately logarithmic) dependence on $\beta$ as can be seen in Fig.~\ref{fig:Example_betaDependence}.
\end{rem}

\begin{rem} \label{rem:numberSamplesComplexity}
	In contrast to the scenario approach~\cite{Calafiore2006b}, the number of samples does not affect the complexity of the resulting optimization problem, nor the probabilistic guarantees. However, a lower number of samples results in looser bounds on the estimated moments as can be seen in Fig.~\ref{fig:Example_sampleDependence}. This in turn affects the feasibility of the optimization problem.
\end{rem}

\section{Receding Horizon Implementation} \label{sec:recedingHorizonImplementation}
In a receding horizon framework, we account for  dynamics of the obstacle and the new measurements available of its state at each planning stage. We consider a time-varying model for the obstacle so as to incorporate the possibility of nonlinear dynamics through linearization around a nominal trajectory. Our dynamic obstace is thus described as
\begin{subequations} \label{eq:dynamicObstacleDynamics}
\begin{align}
	\chi_{t+1} & = E_t \chi_{t} + F_t + w_t \,, \label{eq:dynamicObstacleProcess} \\
	y_t & = H_t \chi_t + v_t \,, \label{eq:dynamicObstacleMeasurement}
\end{align}
\end{subequations}
where $\chi_t \in \Reals^{n_\chi}$ is the obstacle's state, $\chi_0 \sim \mathcal{N}(\mu_{\chi_0},\Sigma_{\chi_0})$ is the initial state, $y_t \in \Reals^{n_y}$ is the measurement and $w_t \sim \mathcal{N}(0,\Sigma_{w_t}),v_t \sim \mathcal{N}(0,\Sigma_{v_t})$ are white noise signals. The obstacle is represented as a polyhedron with $G$ faces, with each face described by:
\begin{equation} \label{eq:dynamicObstacleFaces}
	\{ x \, | \, {a^t_i}^\Trans x + {c^t_i}^\Trans \chi_t + d_i = 0  \}\,,
\end{equation}
where  $ t \in \Nats[1,N],  i \in \Nats[1,G]$, the constants $a^t_i$, $c^t_i$ and $d_i$ are known based on the obstacle's shape. As such, the constraint that $x_t$ must remain outside of the moving obstacle is written as:
\begin{equation} \label{eq:dynamicObstacleRegion}
	\overline{\mathcal{O}^t} \coloneqq \left\{ x_t \in \Reals^{n_x} : \vee_{i=1}^{G}  {a^t_i}^\Trans x_t + c_i^\Trans \chi_t + d_i > 0 \right\} \,.
\end{equation}
This type of constraint is the same as~\eqref{eq:obstacleRegionComplement}, with ${a^t_{ij}}$ now deterministic instead of uncertain and $b^t_{ij}$ capturing the uncertainty of the dynamic obstacle. 

\subsection{Estimation and prediction of obstacle's motion}
Given the linear dynamics and the estimates of the uncertainties, we use a Kalman filter to estimate position of the obstacle ~\cite{Kalman1960}. Let $\hat{\chi}_{t|t-1}$ and $\hat{\chi}_{t|t}$ be the a priori and a posteriori state estimates at time $t$ given measurements up to and including time $t-1$ and $t$, respectively,with $\hat{\Sigma}_{t|t-1}$ and $\hat{\Sigma}_{t|t}$  the corresponding error covariances.  The prediction step of the Kalman filter is:
\begin{subequations} \label{eq:kalmanFilterPriorUpdate}
\begin{align}
	\hat{\chi}_{t|t-1} & = E_{t-1} \hat{\chi}_{t-1|t-1} + F_{t-1} \,, \label{eq:kalmanFilterPriorUpdateMean} \\
	\hat{\Sigma}_{t|t-1} & = E_{t-1} \hat{\Sigma}_{t-1|t-1} {E_{t-1}}^\Trans + \Sigma_{w_{t-1}} \label{eq:kalmanFilterPriorUpdateVariance} \,,
\end{align}
\end{subequations}
with the initialization $\hat{\chi}_{0|0} = \mu_{\chi_0}$ and $\hat{\Sigma}_{0|0} = \Sigma_{\chi_0}$. The update step of the Kalman filter is:
\begin{subequations} \label{eq:kalmanFilterPosteriorUpdate}
\begin{align}
	K_{t} & = \hat{\Sigma}_{t|t-1} + H_t^\Trans \left( \Sigma_{v_t} + H_t \hat{\Sigma}_{t|t-1} H_t^\Trans \right)^{-1} \,, \label{eq:kalmanFilterPosteriorUpdateGain} \\
	\hat{\chi}_{t|t} & = \hat{\chi}_{t|t-1} + K_{t} \left( y_t - H_t \hat{\chi}_{t|t-1} \right) \,, \label{eq:kalmanFilterPosteriorUpdateMean} \\
	\hat{\Sigma}_{t|t} & = \left( I - K_t H_t \right) \hat{\Sigma}_{t|t-1} \,. \label{eq:kalmanFilterPosteriorUpdateVariance}
\end{align}
\end{subequations}

In order to plan safe trajectories we need to predict the future positions of the obstacle with quantifiable confidence. Although this prediction could be done by propagating the estimates $\hat{\chi}_{\tau|\tau}, \hat{\Sigma}_{\tau|\tau}$ forward using~\eqref{eq:kalmanFilterPriorUpdateMean},~\eqref{eq:kalmanFilterPriorUpdateVariance}, in practice the moments of $w_t$ will not be known exactly and thus we cannot quantify and minimize the probability of constraint violation. To this end, we utilize past sample data to get a confidence bound on the predicted trajectory of the obstacle.

Specifically, we assume that we have $N_s$ samples of $w_t$ available for the next $N$ steps of the planning horizon, i.e.\ samples $w^{(1)}_t, \dots, w^{(N_s)}_t$ for all $t \in \Nats[1,N]$, alternatively denoted as $\{ w_t^{(i)} \}_{i=1}^{N_s}$. The availability of these samples is justified either through the use of a generative model (as done for example in~\cite[Section~IV]{Lefkopoulos2019}) or by considering that the obstacle maneuvers are usually extracted through a clustering algorithm based on previously collected sample data~\cite{Lefevre2014,Carvalho2016}. In both cases a set of trajectory data $\{ \chi_t^{(i)} \}_{i=1}^{N_s}$ is  available, from which the model uncertainty $\{ w_t^{(i)} \}_{i=1}^{N_s}$ can be calculated by comparison with the nominal response of linear model~\eqref{eq:dynamicObstacleProcess}. We use these samples directly to forward simulate the process model~\eqref{eq:dynamicObstacleProcess} $N$ steps and to estimate the uncertainty moments required in the Kalman filter equations above. Consequently, we can form the sample estimates~\eqref{eq:sampleMeanEstimate} and~\eqref{eq:sampleCovarianceEstimate} of the obstacle's predicted state. The results of Section~\ref{sec:momentsRobustApproach} can then be directly used to robustify against the uncertainty of the dynamic obstacle's future trajectory. This approach is part of the complete closed-loop algorithm that will be formally presented below.

\subsection{Closed loop}
Given a planning horizon $N$, let $u_{t-1|\tau}$ and $x_{t|\tau}$, with $t \in \Nats[\tau+1,\tau+N]$, denote the optimization problem solutions at time $\tau$. The Chance-Constrained Receding Horizon (CCRH) is presented in Algorithm~\ref{alg:CCRHOC}.
\begin{algorithm}[H]
	\caption{Chance-Constrained Receding Horizon}
	\label{alg:CCRHOC}
	\begin{algorithmic}[1] 
	    \State Given samples $\{ w_t^{(i)} \}_{i=1}^{N_s}$ for $t \in \Nats[0,2N-1]$ from past data or generative model
		\For{$\tau = 0$ to $N-1$}
    		\If{$\tau = 0$}
    		    \State initialize $\hat{\chi}_{0|0}, \hat{\Sigma}_{0|0}$ to $\mu_{\chi_0}, \Sigma_{\chi_0}$
    		\Else
    		    \State obtain measurement $y_{\tau}$ from~\eqref{eq:dynamicObstacleMeasurement}
    		    \State calculate $\hat{\chi}_{\tau|\tau}, \hat{\Sigma}_{\tau|\tau}$ from~\eqref{eq:kalmanFilterPosteriorUpdateMean}
    		\EndIf
    		\State propagate samples $\{ \chi_{\tau}^{(i)} \}_{i=1}^{N_s}$ using~$\{ w_t^{(i)} \}_{i=1}^{N_s}$ and~\eqref{eq:dynamicObstacleProcess} to get $\{ \chi_{t}^{(i)} \}_{i=1}^{N_s}$ for $t \in \Nats[\tau+1,\tau+N]$
    		\State calculate sample moments $\hat{\chi}_{t|\tau}, \hat{\Sigma}_{t|\tau}$ using~\eqref{eq:sampleMeanEstimate},~\eqref{eq:sampleCovarianceEstimate} for $t \in \Nats[\tau+1,\tau+N]$
    		\State measure $x_{\tau}$ and solve Problem~\eqref{eq:mraProblem} using~\eqref{eq:improvedRiskAllocation}
    		\State apply first input $u_{\tau|\tau}$
		\EndFor
	\end{algorithmic}
\end{algorithm}

\begin{assum} \label{ass:CCRHOC_feasibility}
	Each of the $N$ optimization problems of Algorithm~\ref{alg:CCRHOC} are feasible.
\end{assum}

\begin{thm} \label{thm:CCRHOC_safety}
	The sequence of states $x_{1|0},x_{2|1},\dots,x_{N|N-1}$ that is obtained as a result of Algorithm~\ref{alg:CCRHOC} satisfies:
	\begin{equation} \label{eq:CCRHOC_safety}
		\Pr(\wedge_{t=1}^{N} x_{t|t-1} \in \mathcal{X}_t) \geq 1-\epsilon \,,
	\end{equation}
	with a probability according to Theorem~\ref{thm:mraProblemFeasibility}.
\end{thm}

\begin{pf}
	In Algorithm \ref{alg:CCRHOC} every problem is solved using the risk allocation~\eqref{eq:improvedRiskAllocation}, which implies that:
	\begin{equation} \label{eq:stagewiseSafety}
		\Pr(x_{\tau+t|\tau} \in \mathcal{X}_{\tau+t}) \geq 1 - \frac{\epsilon}{N}, \quad \forall t \in \Nats[1,N] \,,
	\end{equation}
	holds for any $\tau \in\Nats[0,N-1]$. Summing up the probabilities of the complement events of~\eqref{eq:stagewiseSafety} we get:
	\begin{equation} \label{eq:horizonSafetyComplement}
		\sum_{\tau=0}^{N-1} \Pr(x_{\tau+1|\tau} \notin \mathcal{X}_{\tau+1}) \leq \epsilon \,.
	\end{equation}
	Using Boole's inequality and~\eqref{eq:horizonSafetyComplement} we obtain:
	\begin{equation}
		\Pr(\vee_{\tau=0}^{N-1} x_{\tau+1|\tau} \notin \mathcal{X}_{\tau+1}) \leq \epsilon \,,
	\end{equation}
	from the complement of which the statement of the theorem follows.
\end{pf}

\section{Simulations} \label{sec:simulations}
We will present two simulations\footnote{The code used for the simulations is available on \href{https://github.com/Exomag/finite-sample-uncertainty-planning-automatica}{\tt finite-sample-uncertainty-planning-automatica}.} to demonstrate the applicability and effectiveness of our approach. The first, illustrates approach of Section~\ref{sec:momentsRobustApproach} on a  simple scenario, to present the safety and optimality of the method, in comparison with other potential approaches and as a function of the number of samples. The second case study, in Section~\ref{sec:closedLoop} illustrates the approach of Section~\ref{sec:recedingHorizonImplementation} on a more realistic case study. All computations were carried out on an Intel i5 CPU at \SI{2.50}{\GHz} with \SI{8}{\giga\byte} of memory using YALMIP~\cite{Lofberg2004} and CPLEX~\cite{CPLEX}.

\subsection{Open-loop case study} \label{sec:openLoop}
We consider a robot with the objective to reach a desired target position while avoiding two \emph{uncertain} walls that block its path. The robot motion model is:
\begin{equation} \label{eq:Study1_EgoDynamics}
	\begin{bmatrix}
		\dot{x}_1 &
		\dot{x}_2
	\end{bmatrix}^\Trans
	= \begin{bmatrix}
		u_1 &
		u_2
	\end{bmatrix}^\Trans \,,
\end{equation}
where the state $x \in \Reals^2$ is its position $(x_1,x_2)$ and is constrained to the set $ \mathcal{X} \coloneqq \{ x \in \Reals^2 : 0 \leq x_i \leq 9, i = \Nats[1,2] \} $. The velocities $(u_1,u_2)$ are the control inputs and are constrained to the set $ \mathcal{U} \coloneqq \{ u \in \Reals^2 : \norm*{u}_\infty \leq 1 \} $. The objective is to minimize the distance between the robot's position and the target position $x_d = [8,7]^\Trans$, encoded through the cost $J = \norm*{x_N-x_d}^2_2$, and the initial state is $x_0 = [1,1]^\Trans$.

The coefficients describing the two walls $d^t_1, d^t_2 \in \Reals^3$ are uncertain and at time $t$ are distributed as:
\begin{subequations} \label{eq:Study1_CoefficientDistribution}
	\begin{align}
	d^t_1 & \sim \Gauss([-1,0,2]^\Trans, 0.001 I_3) \,, \label{eq:Study1_CoefficientDistribution1} \\
	d^t_2 & \sim \Gauss([0,1,6]^\Trans, 0.001 I_3) \,, \label{eq:Study1_CoefficientDistribution2}
	\end{align}
\end{subequations}
for all $t \in \Nats[1,N]$. In order to deal with the uncertain walls, we enforce the joint chance constraint:
\begin{equation} \label{eq:Study1_ChanceConstraint}
	\Pr(\wedge_{t=1}^{N} \vee_{i=1}^{2} [x_{1,t} , x_{2,t} , 1] d^t_{i} > 0) \geq 1 - \epsilon \,.
\end{equation}

We consider the moments of the distributions~\eqref{eq:Study1_CoefficientDistribution1} and~\eqref{eq:Study1_CoefficientDistribution2} to be \emph{unknown}, and thus draw $N_s$ i.i.d.\ samples from each one and reformulate the resulting problem as per the Moments Robust Approach (MRA) of Section~\ref{sec:momentsRobustApproach}. Such samples could originate from online measurement data (e.g.\ through the use of a LIDAR unit) that are collected from the environment, such that trajectory planning can subsequently commence. For comparison reasons we also solve the problem assuming perfect knowledge of the moments, henceforth called \enquote{Exact Moments Approach} (EMA), and using the Scenario Approach (SA).

The dynamics~\eqref{eq:Study1_EgoDynamics} are discretized with sampling time $T_s = \SI{1}{\second}$ for a planning horizon of $N = 10$ (i.e.\ $\SI{10}{\second}$).  The joint chance constraint~\eqref{eq:Study1_ChanceConstraint} is imposed with $\epsilon = \num{0.05}$ and $\beta = \num{e-3}$ for the whole horizon. The number of samples used is $N_s^{\text{MRA}}=N_s^{\text{SA}}=\num{1259}$\footnote{The number of samples $N_s^{\text{SA}}$ was chosen such that the SA would provide the safety and certainty guarantees prescribed~\cite{Sessa2018}. The number of samples $N_s^{\text{MRA}}$ was chosen to be the same for comparison reasons.}. The resulting optimization problem is a MISOCP with $\num{40}$ continuous variables, $\num{20}$ binary variables and $\num{132}$ constraints in the case of the MRA/EMA and a MIQP\footnote{Mixed-Integer Quadratic Program} with $\num{40}$ continuous variables, $\num{20}$ binary variables and $\num{25292}$ constraints in the case of the SA. The problem is solved with all methods $\num{100}$ different times, with different realizations of the disturbance. The risk distribution iterative algorithm of~\cite[Section~5]{Jha2018} is used for the MRA and the EMA to improve the risk allocation with a small increase in computation time.

\begin{figure}[t!]
	\centering
	\setlength{\figureheight}{0.4\columnwidth}
	\setlength{\figurewidth}{0.4\columnwidth}
	\includegraphics{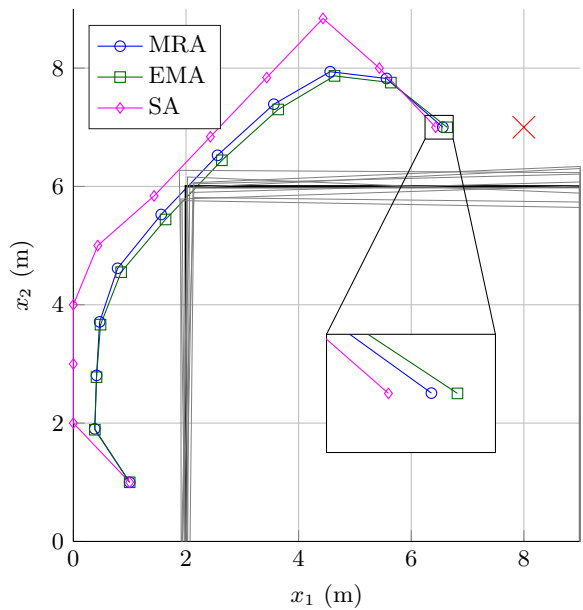}
	\caption{One simulation of the robot's trajectories for the MRA (blue circles), EMA (green squares) and SA (magenta diamonds). The expected positions of the two uncertain walls are displayed (black) along with 10 exemplary wall samples (gray) and the terminal target position (red cross).}
	\label{fig:OpenLoop_Trajectories}
\end{figure}

The solutions to one instance of the problem are presented in Fig.~\ref{fig:OpenLoop_Trajectories}. Under moment uncertainty (MRA) the robot chooses a slightly wider maneuver compared to perfect moment knowledge (EMA), in order to avoid the uncertain position of the walls, and has a final position slightly farther away from the target position. The SA on the other hand performs a wider maneuver and ends up even farther away from the desired target. As expected, robustifying against moment uncertainty using the MRA produced a slightly worse solution, to account for the increased wall uncertainty compared to the EMA.

We evaluated the empirical violation probability of each method through Monte Carlo simulations using $\num{e+5}$ new realizations of the walls' unknown positions. The distribution of the violation probability and the total cost over the $\num{100}$ different instances solved is presented in Fig.~\ref{fig:OpenLoop_CostProbabilityBoxplot}. As seen in Fig.~\ref{fig:OpenLoop_CostBoxplot}, the EMA produced the smallest cost (consistently, since there  is no stochasticity involved in its solution), with the MRA having the next smaller one and the SA having the largest cost. As seen in Fig.~\ref{fig:OpenLoop_ProbabilityBoxplot} the MRA had a violation probability of approximately \SIrange{1}{2}{\percent}, the EMA approximately \SI{4.8}{\percent}, and the SA was the most conservative with a probability of \SIrange{0}{1}{\percent}, with all methods having a violation probability within the prescribed safety margin of $\SI{5}{\percent}$.

We evaluated the total cost and the solver time for a varying number of samples $N_s$ where for each value of $N_s$ the cost/time values were averaged over 10 different realizations of the problem. The total cost of each method is presented in Fig.~\ref{fig:OpenLoop_Costs}. The MRA cost converges to the EMA cost as $N_s$ increases (cf.\ Theorem~\ref{thm:mraProblemFeasibility}). The cost of the SA on the other hand increases with $N_s$, since the solution of the SA converges to the robust counterpart of Problem~\eqref{eq:initialProblem} as $N_s \rightarrow \infty$. The solver time of each method is presented in Fig.~\ref{fig:OpenLoop_SolveTimes}. The MRA and EMA solver times stay approximately constant as $N_s$ increases, whereas the SA solver time increases since a higher number of samples leads to a higher number of constraints.

\subsection{Closed-loop case study} \label{sec:closedLoop}
In this case study we examine a closed-loop trajectory planning scenario. The controlled car (henceforth called \enquote{ego car}) is driving with an initial forward velocity of \SI{19.44}{\meter\per\second} on the left/fast lane of a two-lane highway when an uncontrolled car (henceforth called \enquote{adversary car}) driving on the right/slow lane starts merging in front of the ego car with an unchanging orientation, under the assumption of a slow maneuver.

We model the ego car dynamics as a double integrator:
\begin{equation} \label{eq:Study2_EgoDynamics}
	\begin{bmatrix}
		\dot{x}_1 \\
		\dot{x}_2 \\
		\dot{x}_3 \\
		\dot{x}_4
	\end{bmatrix}
	= \underbrace{\begin{bmatrix}
		0 & 0 & 1 & 0 \\
		0 & 0 & 0 & 1 \\
		0 & 0 & 0 & 0 \\
		0 & 0 & 0 & 0
	\end{bmatrix}}_{\coloneqq A}
	\begin{bmatrix}
		x_1 \\
		x_2 \\
		x_3 \\
		x_4
	\end{bmatrix}
	+ \underbrace{\begin{bmatrix}
		0 & 0 \\
		0 & 0 \\
		1 & 0 \\
		0 & 1
	\end{bmatrix}}_{\coloneqq B}
	\begin{bmatrix}
		u_1 \\
		u_2
	\end{bmatrix} \,,
\end{equation}
where the state $x \in \Reals^4$ contains the two-dimensional position $(x_1,x_2)$ of the car and the corresponding velocities $(x_3,x_4)$. The accelerations $(u_1,u_2)$ are the control inputs and are constrained to the set $\mathcal{U} \coloneqq \{ u \in \Reals^2 : \abs*{u_1} \leq \SI{10}{\meter\per\second} , \abs*{u_2} \leq \SI{2}{\meter\per\second} \}$. The objective is to minimize the deviation of the car's position from the center of the left lane and its velocity from \SI{19.44}{\meter\per\second}, while also minimizing the use of the control inputs. As such, the cost is encoded as:
\begin{equation}
	J = \sum_{t=1}^{N} \norm{\begin{bmatrix}x_{t,2} \\ x_{t,3}\end{bmatrix} - \begin{bmatrix}2 \\ 19.44\end{bmatrix}}^2_2 + 0.1 \norm{\begin{bmatrix}u_{t-1,1} \\ u_{t-1,2}\end{bmatrix}}^2_2\,.
\end{equation}
The dynamics~\eqref{eq:Study2_EgoDynamics} are discretized with sampling time $T_s = \SI{0.2}{\second}$ resulting in the discrete form~\eqref{eq:systemDynamicsLTV}. The initial state is $x_0 = [\SI{0}{\meter}, \SI{2}{\meter}, \SI{19.44}{\meter\per\second}, \SI{0}{\meter\per\second}]^\Trans$.

The adversary car dynamics and measurements are:
\begin{subequations}
	\begin{align} 
	    \dot{\chi} & = A \chi + B \nu \,, \label{eq:Study2_AdversaryDynamics} \\
		\begin{bmatrix}
			y_1 \\
			y_2
		\end{bmatrix}
		& = \begin{bmatrix}
			1 & 0 & 0 & 0 \\
			0 & 1 & 0 & 0
		\end{bmatrix}
		\chi \,, \label{eq:Study2_AdversaryMeasurements}
	\end{align}
\end{subequations}
where the state $\chi \in \Reals^4$ contains the position $(\chi_1,\chi_2)$ of the car and the corresponding velocities $(\chi_3,\chi_4)$. The adversary car longitudinal model is detected as one of constant velocity, corresponding to $\nu_1 = 0$. The lateral model is assumed to be a second-order response converging to the center of the left lane, corresponding to $\nu_2 = -[k_1,k_2] [\chi_2,\chi_4]^\Trans$ where the gains are provided by a Linear–Quadratic Regulator (LQR). The dynamics~\eqref{eq:Study2_AdversaryDynamics} are discretized with sampling time $T_s = \SI{0.2}{\second}$, and process/measurement additive noise is introduced, resulting in the discrete form~\eqref{eq:dynamicObstacleDynamics}. The initial state $\chi_0$, process noise $w_t$ and measurement noise $v_t$ are distributed according to:
\begin{subequations}
\begin{align}
	\chi_0 & \sim \Gauss ([0, -2, 19.44, 0]^\Trans, \diag([0, 0, 1.23, 0.08]))  \,, \label{eq:Study2_AdversaryInitialCondition} \\
	w_t & \sim \Gauss (\mymathbb{0}, \diag([0, 0, 0.04, 0.001])) \,, \label{eq:Study2_AdversaryProcessNoise} \\
	v_t & \sim \Gauss (\mymathbb{0}, \diag([1, 0.04])) \,, \label{eq:Study2_AdversaryMeasurementNoise}
\end{align}
\end{subequations}

In order to deal with the uncertain pose $\chi_t$ of the adversary car, we enforce the joint chance constraint:
\begin{equation} \label{eq:Study2_ChanceConstraint}
	\Pr(\wedge_{t=1}^{N} \vee_{i=1}^{4} [ x_{1,t} , x_{2,t} , 1 ] d^t_{i} > 0) \geq 1 - \epsilon \,.
\end{equation}
The exact form of the coefficients $d^t_{i}(\chi_t)$ is:
\begin{subequations}
\begin{align}
	d_1^t & = [ 1 , 0 , - \chi_1 - \flatfrac{L}{2} ]^\Trans , \\
	d_2^t & = [ - 1 , 0 , \chi_1 - \flatfrac{L}{2} ]^\Trans , \\
	d_3^t & = [ 0 , 1 , - \chi_2 - \flatfrac{W}{2} ]^\Trans , \\
	d_4^t & = [ 0 , - 1 , \chi_2 - \flatfrac{W}{2} ]^\Trans ,
\end{align}
\end{subequations}
where $L$ and $W$ are the length and width, respectively, of the adversary car. We consider the moments of the distributions~\eqref{eq:Study2_AdversaryInitialCondition},~\eqref{eq:Study2_AdversaryProcessNoise} and~\eqref{eq:Study2_AdversaryMeasurementNoise} to be \emph{unknown}, and thus draw $N_s$ i.i.d.\ samples from each one and reformulate the resulting problem as per the CCRH approach of Section~\ref{sec:recedingHorizonImplementation}. 

A planning horizon of $N = 25$ (i.e.\ $\SI{5}{\second}$) is chosen. Both cars have length \SI{4}{\meter} and width \SI{2}{\meter}, which are taken into account for the relevant inequality constraints. We also enforce lane constraints. The joint chance constraint~\eqref{eq:Study2_ChanceConstraint} is imposed with $\epsilon = \num{0.05}$ and $\beta = \num{e-3}$ for the whole horizon. The number of samples drawn from the distributions of $x_0$ and is $N_s = 1000$ for every time step; obtainable for example from real-world data of lane change maneuvers of this kind (or a similar generative model). The resulting optimization problem is a MISOCP with $\num{150}$ continuous variables, $\num{100}$ binary variables and $\num{379}$ constraints.

\begin{figure}[t!]
	\centering
	\setlength{\figureheight}{0.5\columnwidth}
	\setlength{\figurewidth}{0.8\columnwidth}
	\includegraphics{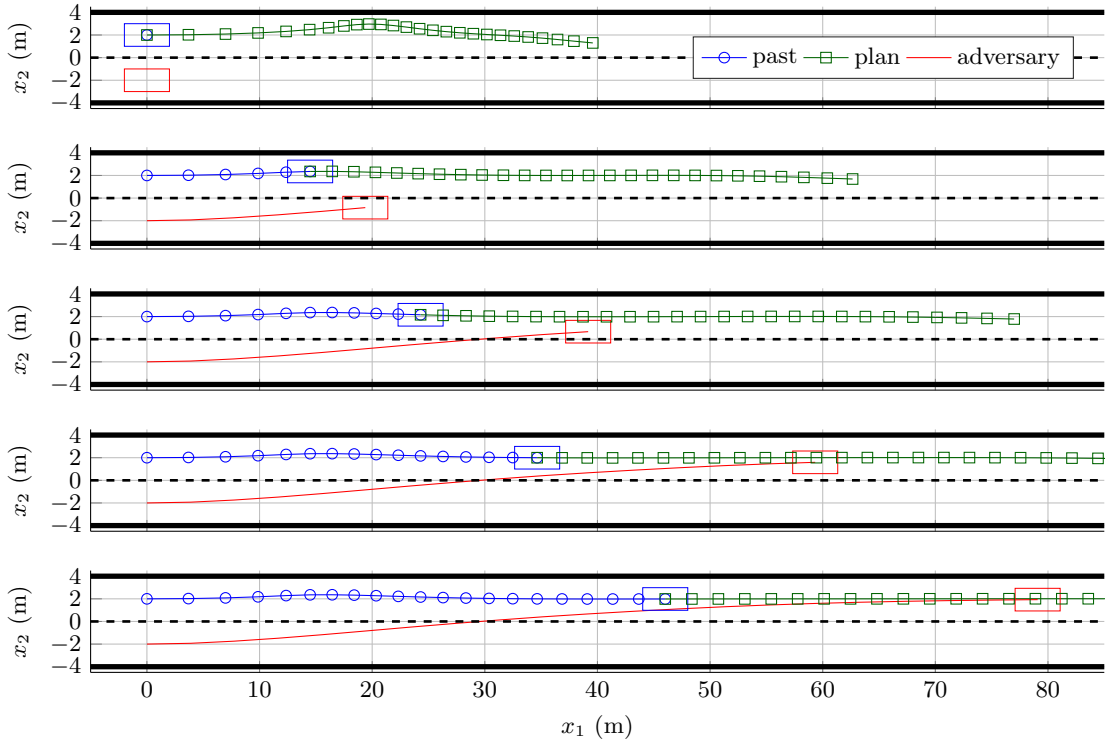}
	\caption{Simulation of the ego car's trajectory. Pictured are time steps $t = \SI{0}{\second}, \SI{1}{\second}, \SI{2}{\second}, \SI{3}{\second}$ and $\SI{4}{\second}$. At every time step the ego car's past closed-loop trajectory is displayed (blue circles), along with the future open-loop plan (green squares) and the adversary car's past trajectory (red).}
	\label{fig:ClosedLoop_Frames}
\end{figure}

The simulation is presented in Fig.~\ref{fig:ClosedLoop_Frames}. The ego car brakes and steers towards the left, in order to avoid the merging adversary car whose position is uncertain. As a result of the initially high estimate uncertainty of the current and future positions of the adversary car, the first open-loop solution calculated (in the first frame of Fig.~\ref{fig:ClosedLoop_Frames}) is to react sharply. In contrast, the closed-loop solution exhibits a less severe and more comfortable maneuver. The more aggressive maneuver of the initial open-loop solution incurs a higher cost of \num{3785.2}, whereas the closed-loop solution has a lower cost of \num{1766.3}. This is also evident in Fig.~\ref{fig:ClosedLoop_Inputs}, where the longitudinal and lateral acceleration inputs of the ego vehicle are displayed. The open-loop trajectory brakes fully both longitudinally and laterally at the start of the horizon for a longer duration than the closed-loop solution.

The improved performance of the closed-loop trajectory is due to the receding horizon implementation's ability to react to the adversary car's latest movement. As more measurements are acquired and passed through the Kalman filter's update steps, the estimates of the adversary's car position improve. In open loop the future state of the adversary car is increasingly more uncertain the farther away in the future we want to estimate, whereas in closed loop the uncertainty settles pretty quickly to a certain confidence (that depends on how strong the process and measurement noise is). This can be observed in Fig.~\ref{fig:ClosedLoop_Uncertainty}, where the longitudinal and lateral velocity estimates of the adversary car are displayed over time. The open-loop estimate uncertainty of the longitudinal velocity grows over time, whereas the closed-loop uncertainty does not.

\section{Conclusion} \label{sec:conclusion}
We tackled the problem of trajectory planning with uncertain polyhedral obstacles using chance-constrained optimization. We reformulated the problem into a deterministic and tractable form based on the uncertainty's moments. We estimated said moments from collected samples, resulting in noisy estimates. We derived tight and asymptotic concentration bounds on said estimates that were used to formulate a robust tractable optimization problem whose solution is feasible with regards to the original problem up to a prescribed confidence bound. We extended this framework to a receding horizon, including obstacles with partially known dynamics. We illustrated both open-loop and closed-loop results through two case studies. Our method produced safe trajectories using only finite sample moment estimates, outperforming previous solutions, with the closed-loop improving over the open-loop results.

\begin{ack}
The authors would like to thank I.\ Usmanova for helping with Lemma~\ref{lem:covarianceBound}.
\end{ack}



\appendix
\section{Additional open-loop case study figures}
\begin{figure}[H]
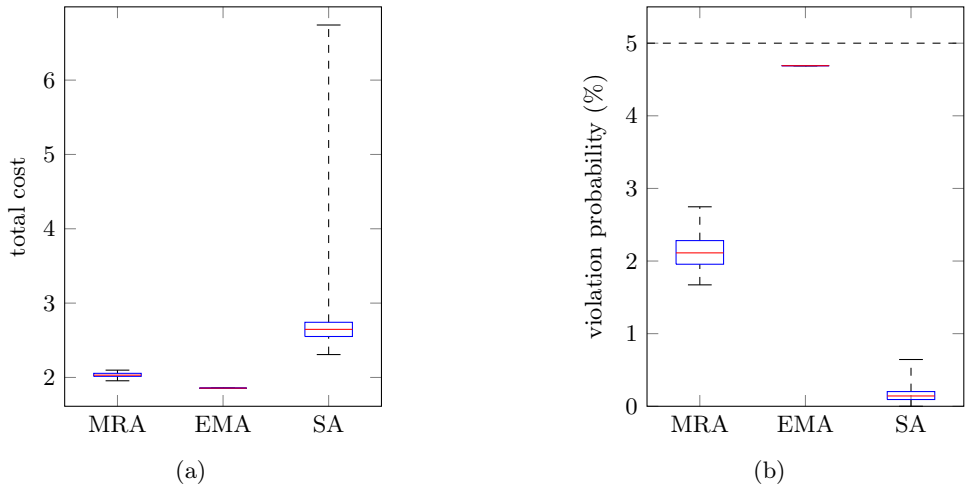

	\centering
	\setlength{\figureheight}{0.3\columnwidth}
	\setlength{\figurewidth}{0.25\columnwidth}
	\subfloat[]{
		\includegraphics{OpenLoop_CostBoxplot.tikz}
		\label{fig:OpenLoop_CostBoxplot}
	}
	\hfil
	\subfloat[]{
		\includegraphics{OpenLoop_ProbabilityBoxplot.tikz}
		\label{fig:OpenLoop_ProbabilityBoxplot}
	}
	\caption{Distribution of the (a) total cost and (b) empirical violation probability over $\num{100}$ instances of the trajectory planning optimization problem. The median is in red and the 25/75-th percentiles are in blue. The black dashed line corresponds to the prescribed safety margin of $\SI{5}{\percent}$.}
	\label{fig:OpenLoop_CostProbabilityBoxplot}
\end{figure}

\begin{figure}[H]
	\centering
	\setlength{\figureheight}{0.3\columnwidth}
	\setlength{\figurewidth}{0.8\columnwidth}
	\includegraphics{OpenLoop_Costs.tikz}
	\caption{Total cost for a varying number of samples $N_s$ for the MRA (blue circles), EMA (green squares) and SA (magenta diamonds). For each value of $N_s$ the cost is averaged over 10 different realizations of the problem.}
	\label{fig:OpenLoop_Costs}
\end{figure}

\begin{figure}[H]
	\centering
	\setlength{\figureheight}{0.3\columnwidth}
	\setlength{\figurewidth}{0.8\columnwidth}
	\includegraphics{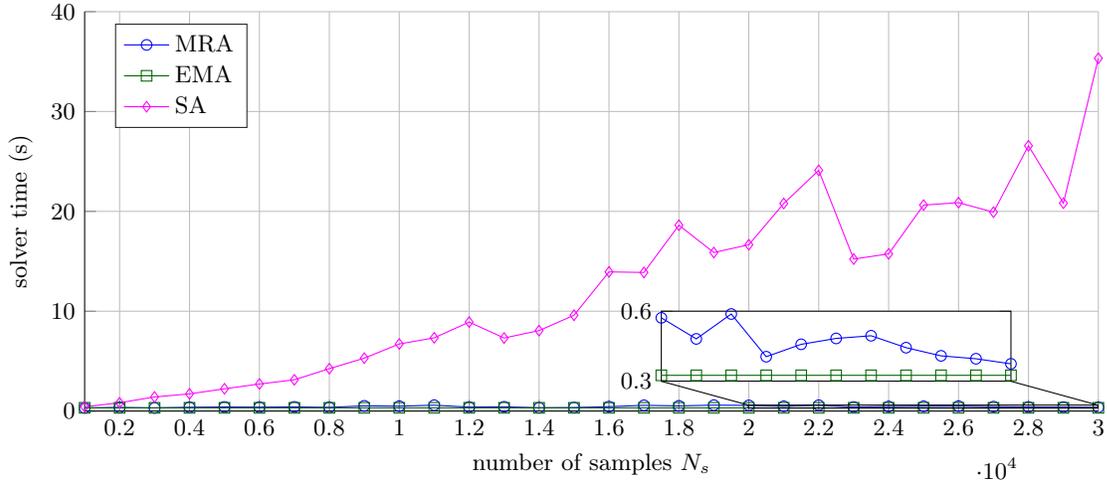}
	\caption{Solver time for a varying number of samples $N_s$ for the MRA (blue circles), EMA (green squares) and SA (magenta diamonds). For each value of $N_s$ the time is averaged over 10 different realizations of the problem.}
	\label{fig:OpenLoop_SolveTimes}
\end{figure}

\section{Additional closed-loop case study figures}
\begin{figure}[H]
	\centering
	\setlength{\figureheight}{0.6\columnwidth}
	\setlength{\figurewidth}{0.8\columnwidth}
	\includegraphics{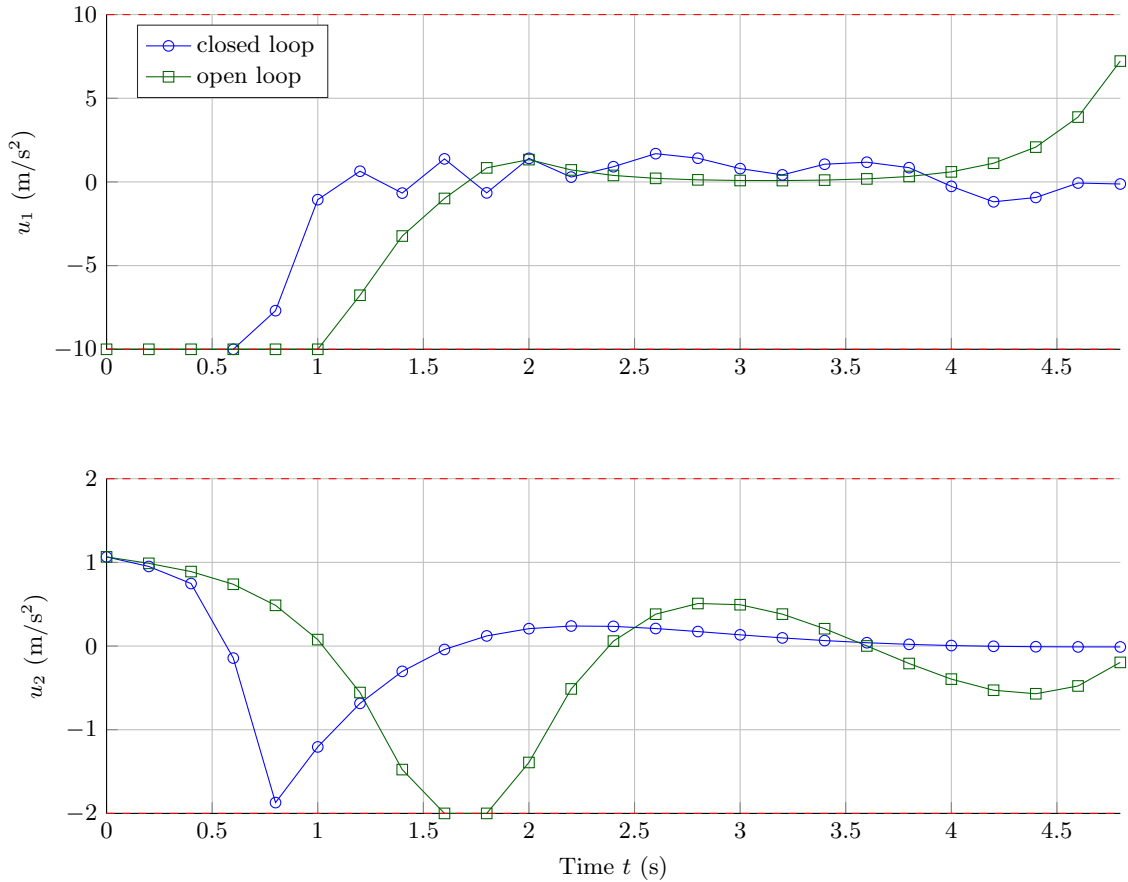}
	\caption{Longitudinal and lateral acceleration inputs of the ego car. The initial open-loop plan is displayed (green squares), along with the inputs actually used during the closed-loop execution (blue circles).}
	\label{fig:ClosedLoop_Inputs}
\end{figure}

\begin{figure}[H]
	\centering
	\setlength{\figureheight}{0.6\columnwidth}
	\setlength{\figurewidth}{0.8\columnwidth}
	\includegraphics{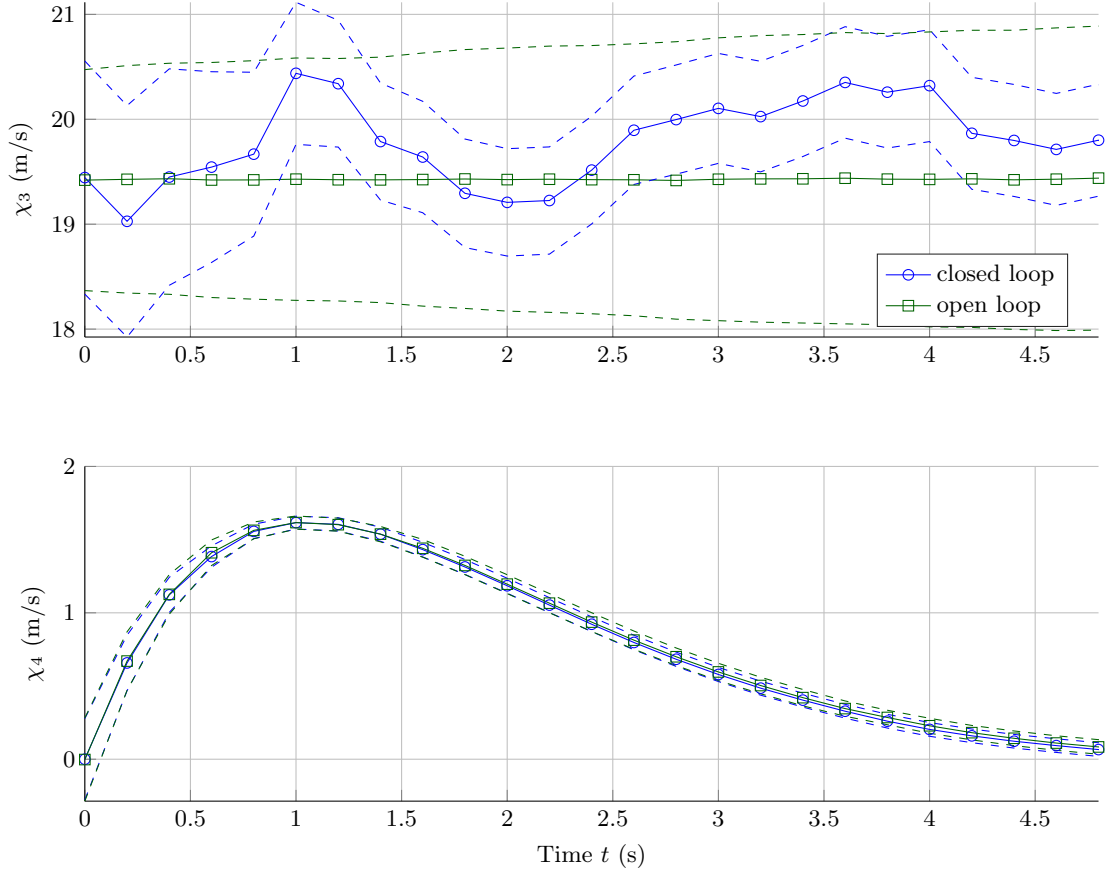}
	\caption{Longitudinal and lateral velocity estimates of the adversary car. The initial open-loop estimates are displayed (green squares), along with the estimates produced during the closed-loop execution (blue circles) and the $\pm 1$ standard deviation interval (blue/green dashed lines).}
	\label{fig:ClosedLoop_Uncertainty}
\end{figure}

\end{document}